\documentclass[pdflatex,sn-nature]{sn-jnl}

\usepackage{graphicx}%
\usepackage{multirow}%
\usepackage{amsmath,amssymb,amsfonts}%
\usepackage{amsthm}%
\usepackage{mathrsfs}%
\usepackage[title]{appendix}%
\usepackage{xcolor}%
\usepackage{textcomp}%
\usepackage{manyfoot}%
\usepackage{booktabs}%
\usepackage{algorithm}%
\usepackage{algorithmicx}%
\usepackage{algpseudocode}%
\usepackage{listings}%
\usepackage{siunitx}

\usepackage{subcaption}
\usepackage{caption}
\usepackage{colortbl}
\usepackage{tabularx}
\usepackage{tikz}
\usepackage{changepage}

\newcommand{\actantgrid}[6]{%
\begin{tikzpicture}[baseline=0.6ex]
  \def\s{0.2} 
  \def\w{0.4} 
  \fill[#1] (0,\s) rectangle (\w,2*\s);
  \fill[#2] (\w,\s) rectangle (2*\w,2*\s);
  \fill[#3] (2*\w,\s) rectangle (3*\w,2*\s);
  \fill[#4] (0,0) rectangle (\w,\s);
  \fill[#5] (\w,0) rectangle (2*\w,\s);
  \fill[#6] (2*\w,0) rectangle (3*\w,\s);
  \draw (0,0) grid[xstep=\w, ystep=\s] (3*\w,2*\s);
\end{tikzpicture}%
}

\theoremstyle{thmstyleone}%
\theoremstyle{thmstyletwo}%

\theoremstyle{thmstylethree}%

\begin{document}

\title[Article Title]{Measuring Narrative Polarization in Online Discourse}

\author[1]{\fnm{Jan} \sur{Elfes}}\email{jan.elfes@ucd.ie}

\author[1]{\fnm{Marco T.} \sur{Bastos}}\email{marco.bastos@ucd.ie}

\author*[2,3]{\fnm{Luca Maria} \sur{Aiello}}\email{luai@itu.dk}

\affil[1]{\orgdiv{Information and Communication Studies}, \orgname{University College Dublin}, \orgaddress{\country{Ireland}}}

\affil[2]{\orgdiv{Data Science}, \orgname{IT University of Copenhagen}, \orgaddress{\country{Denmark}}}

\affil[3]{\orgname{Pioneer Centre of AI}, \orgaddress{\country{Denmark}}}

\abstract{Polarization research has demonstrated how people cluster in homogeneous groups with opposing opinions. However, this effect emerges not only through interaction between people, limiting communication between groups, but also between narratives, shaping opinions and partisan identities. Yet, how polarized information environments portray opposing interpretations of reality, and whether narratives move between content environments despite limited interactions, remains unexplored. To address this gap, we formalize the concept of \textit{narrative polarization} and demonstrate its measurement in 212 YouTube videos and 90,029 comments on the Israeli-Palestinian conflict. Based on structural narrative theory and implemented through a large language model,  we extract the narrative roles assigned to central actors in two partisan information environments. We find that while videos produce highly polarized narratives, comments exhibit significantly lower levels of narrative polarization, converging on shared core narrative structures. However, recurring \textit{narrative motifs} capturing more complex actor constellations reveal additional differences between partisan environments.}


\maketitle

\section{Introduction}
Many societies—particularly Western democracies—appear increasingly fractured along political and cultural lines, eroding the foundations of democratic deliberation~\cite{mccoy_polarization_2018, iyengar_origins_2019, jacoby_is_2014, bolet_drinking_2021}. Social media platforms have emerged as a crucial infrastructure in this process, simultaneously mobilizing followers and reshaping political narratives around these emerging divides~\cite{yasseri_political_2016, castelli_gattinara_far_2019, gilardi_social_2022}. Understanding the process by which individual differences turn into polarized societies requires attending to the rhetorical dimension of how collective narratives emerge.

Polarization generally refers to the divergence of beliefs across different groups. This divergence can take various shapes, captured by a diverse range of polarization measures~\cite{bramson_understanding_2017, martino_quantifying_2025}. Further, different types of polarization have been observed, including affective polarization, related to negative perceptions of an out-group, and ideological polarization, which describes divergent opinions on a specific issue~\cite{lelkes_mass_2016}. What unites these different measures and types is that they rely on measuring people's attitudes, usually through surveys or behavior-based methods~\cite{kish_bar-on_unraveling_2024}. Social media data has facilitated the inference of opinion distributions from user interactions at a large scale. Moreover, it revealed the structures that undergird online polarization dynamics. Research has found that both algorithmic content curation and self-selection steer users towards more partisan content and information sources~\cite{cinelli_selective_2020, robertson_users_2023, ekstrom_search_2024, pournaki_how_2025}, resulting in ideologically homogeneous clusters with limited cross-cutting communication~\cite{bakshy_exposure_2015, bessi_users_2016, flamino_political_2023}. Within these \textit{echo chambers}, polarization intensifies through exposure to aligned viewpoints~\cite{sunstein_law_2002}. Across these studies, user attitudes are commonly inferred from interactions such as likes, shares, and comments with prominent partisan accounts or content~\cite{falkenberg_towards_2025}. These surface-level interactions are sometimes complemented by textual analysis, for example, identifying polarizing language, sentiments, or topics~\cite{simchon_troll_2022, nemeth_scoping_2023, karjus_evolving_2024}. However, this predominantly focuses on linguistic divergence between groups, serving as a signal of group membership. It does not capture the underlying argumentative and rhetorical mechanisms that drive polarization.

In contrast to this body of work, we understand narratives as abstract patterns that establish ``meaningful correlations between events [and] offer the potential for identification and interpretation to community members''~\cite[p. 7]{baier_narratives_2023}. They are fundamental to both the formation of individual and collective identities and political deliberation~\cite{somers_narrative_1994, bliuc_cooperation_2022}. Narratives have been linked to the long-term persuasiveness of messages, and experimental evidence suggests that the mere exposure to polarized narratives affects individual beliefs~\cite{oschatz_long-term_2020, antinyan_narratives_2024}. Hence, understanding how polarization emerges online requires examining the narratives around contested issues and their distribution across partisan information environments. Facilitated by advances in computational methods, empirical work has only recently begun to examine online polarization through this narrative lens. Initial evidence suggests that polarized user groups advance distinct narratives and that certain narrative divides, chiefly between mainstream and non-mainstream framings, align users across ostensibly disparate topics~\cite{loru_ideology_2025, pournaki_conflicting_2025}. 

We introduce the concept of \textit{narrative polarization} to measure diverging representations of key actors through their positioning in popular narratives. We formalize narrative through narratological theory. Chiefly, we utilize Greimas' Actantial Model~\cite{greimas_structural_1984} to capture underlying narrative assumptions of a text through a set of six character roles, called \textit{actants}. Thus, narrative polarization emerges as structural differences in actant constellations between partisan information environments. 

To test this concept, we investigate the distribution of narratives around the Israeli-Palestinian conflict on YouTube, a prominent case study of how narratives shape partisan identities and perpetuate conflicts~\cite{bar-tal_sociopsychological_2014}. The October 7, 2023 terror attack by Hamas and Israel's subsequent military invasion of Gaza have spurred international attention, transforming support for different factions into a highly polarized issue in public debate. Much of the discussion takes place online, where social media platforms shape the discourse. YouTube in particular serves as a news source for an increasing number of users~\cite{newman_reuters_2025}, while its affordances for commenting enable users to reactively shape the discourse around an issue.

We design two sets of partisan search queries. Each focuses on issues stressed by one specific faction, for example, antisemitism in the Israeli-leaning discourse or Zionism in the Palestinian-leaning discourse. From each set, we compose a partisan information environment, including the most relevant videos from the YouTube recommender algorithm and their comments. Due to the partisan issue focus of the queries, we expect the different sets of videos to propagate diverging narratives about the conflict. Furthermore, we are interested in the extent to which comments interact with this effect. Given that comments are less influenced by algorithmic decisions, we investigate whether they provide counterpoints to a biased representation in the videos or whether they reinforce the initial query slant. We find that the videos propagate different narratives about the conflict. Further, although comments generally follow the agenda set by the videos, they exhibit lower divergence of core narratives, offering a more unified perspective across partisan information environments.

This disconnect between videos and comments raises a critical interpretive question: do partisan information environments contain polarized narratives, portraying the same actors and events through incompatible ideological frameworks? Or do they simply emphasize different aspects of the conflict without opposing representations? To test this, we center the analysis on the main sides of the conflict, that is, Israeli and Palestinian actors, and measure how different objects, such as violence or rights, are attributed differently to these subjects across the query groups. This \textit{subject divergence} reveals whether the different environments portray the interests and roles of central actors differently, rather than just focusing on different aspects of the conflict. The results confirm that narrative differences between the environments stem from opposing representations of the central conflict parties. The videos diverge notably in their portrayal of which side pursues violence and which has security interests. In contrast, comments exhibit significantly lower subject divergence between partisan environments.

However, this trend primarily refers to core narratives that capture assumptions about the central narrative function of desire, namely, who pursues what. Considering additional functions such as control and opposition, and their respective actants, we identify a set of \textit{narrative motifs} that further differentiate representations of actors in the two partisan environments. These specific role constellations capture additional dimensions, such as internal actor complexity, justification of violence, and actor dependencies. Consequently, while comments offset the initial bias of videos on core narratives, fundamentally different narrative representations persist across more complex actor constellations. This suggests that narrative polarization manifests at various levels of complexity.

\section{A structural model of narrative}
\begin{figure}
    \centering
    \includegraphics[width=1\linewidth]{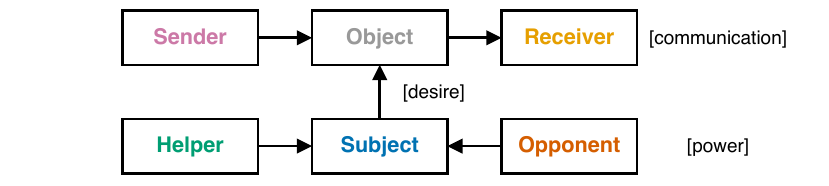}
    \caption{\textbf{Greimas' Actantial Model}. The Actantial Model~\cite{greimas_structural_1984} contains six actants. These functional characters are arranged along three axes: desire, communication, and power. Arrows indicate the direction of the relationship. A minimal example describes a prince (Subject) who desires land (Object), which the king (Sender) will bestow upon him (Receiver), should he overcome the dark knight (Opponent), aided by a magical companion (Helper).} 
    \label{fig:actantial_model}
\end{figure}

Structuralist narratology sought to provide clear language and frameworks for analyzing narrative features of text. This formal structure proves beneficial in computational narrative analysis. Structural models help delineate clear features for automated extraction and analysis. These features allow analysis to separate narrative aspects from other semantic features of the text. Moreover, they provide a consistent representation that is comparable across their applications to different sources and contexts. The structured approach offers a bridge between the complex, nuanced concept of narrative and the precise requirements of computational application. We employ Greimas' Actantial Model~\cite{greimas_structural_1984} as a scaffolding to explore the narrative nuances within our data (Fig. \ref{fig:actantial_model}).

The model comprises six functional categories structured by three fundamental relationships. The Subject desires the Object, which is in turn communicated from the Sender to the Receiver. Additionally, the Helper and the Opponent exert power on the Subject to influence its quest to obtain the Object. The positioning of actors within this model reveals underlying assumptions. The Object establishes the text's main focus, while the Subject determines whose interests and actions are foregrounded. The Sender controls the Object and thus holds significant power and agency, whereas the Receiver is the observed or intended target. Finally, the Helper and Opponent relate directly to the Subject as supporter or adversary, respectively. The framework was originally developed on folktales and myths, for example, describing the desire of a prince (Subject) to obtain land (Object), which is controlled by the king (Sender) and given to the prince (Receiver) if he defeats a dark knight (Opponent), with the help of his magical sidekick (Helper). However, given the general form of the model, this structure directly translates to contemporary issues, for example, by outlining the Israeli desire for security, provided by their military against the opposition of Hamas, or Palestinians' desire for land, controlled by Israeli settlers, etc. Crucially, all actors can carry either positive or negative connotations. For instance, when the Subject desires security, the Opponent typically appears as a threat. Conversely, when the Object is violence, the Subject's desire appears negative, rendering opposition closer to resistance. The automated application of this framework using a large language model (DeepSeek-R1-Distill-Qwen-32B) to label the six roles in comments and video transcripts has been extensively tested and validated with human annotations. Details on the methodology and validation results are included in \textit{Materials and Methods}.

\subsection{Narrative polarization}
The Actantial Model is a formalization of narrative deep structure, that is, fundamental assumptions about actors, roles, and relations, that undergird a story~\cite{de_fina_analyzing_2011}. It provides a ground layer, connecting individual events. Applied to YouTube videos and comments, the model abstracts from the individual text, capturing the fundamental assumptions about the conflict. For example, while a comment might describe a specific event, such as the raid on a hospital, the Actantial Model captures whether the perpetrating actor is portrayed as inherently violent or as responding to an existential threat. When aggregated across an information environment, these individual representations coalesce into a general pattern of how conflict actors are portrayed.

Following this, we define \textit{narrative polarization} between two information environments as diverging representations of key actors through their positioning in popular narratives. For example, a representation of Palestinians as a people striving for rights and freedoms and of Israel as a violent actor favors interpretations of violent actions by Palestinians as acts of resistance. In contrast, a focus on the struggle of Jews to escape historical persecution and of Palestinian groups as a constant violent threat might render the same actions as terrorism. Such structural differences in how actors are positioned shape the information landscapes that users navigate.

As such, narrative polarization operates at a different level than issue polarization, which describes the divergence of people's stances on an issue~\cite{bruns_polarization_2024}, and instead concerns the frameworks through which an issue is perceived. This mirrors what Kligler-Vilenchik et al. term interpretive polarization~\cite{kligler-vilenchik_interpretative_2020}. What distinguishes narrative polarization is the application of a narrative lens. Rather than examining surface-level context such as topic distributions or issue framing, we examine the deeper structural representations of actors and their relationships, as formalized through the Actantial Model.

To measure narrative polarization empirically, we introduce \textit{subject divergence} as the difference between two information environments in how an Object is emphasized as the interest or action of Israeli versus Palestinian actors. To operationalize this, we aggregate the IDF, Israel/Israelis, Jews, and Zionists as Israeli actors (IS) and Arabs, Hamas, Muslims, and Palestine/Palestinians as Palestinian actors (PA). For each Object $i$, we compute the share of attributions to IS and PA actors in each environment, where the two shares sum to one, and let $p(i, \mathrm{IS})$ and $q(i, \mathrm{IS})$ denote the IS share in the Israeli-leaning and Palestinian-leaning environments, respectively. We calculate \textit{subject divergence} as their difference:
\begin{equation}\label{eq:sub_div_red}
    D(i) = p(i, \mathrm{IS}) - q(i, \mathrm{IS}).
\end{equation}
The measure ranges from $-1$ to $1$, with values near $0$ indicating low divergence.

Crucially, $D(i)$ is a \textit{relative} measure. It captures how the Israeli-leaning environment differs from the Palestinian-leaning one, not which actor receives more attribution in absolute terms. Consider \textit{violence} ($v$) as an illustrative example: if $p(v, \mathrm{IS})=0.6$ and $q(v, \mathrm{IS})=0.9$ and conversely $p(v, \mathrm{PA})=0.4$ and $q(v, \mathrm{PA})=0.1$, both environments attribute violence predominantly to Israeli actors. However, the Israeli-leaning environment does so \textit{less}. Hence, the resulting difference $D(v)=-0.3$ is labeled as PA-skew as Palestinians receive comparatively greater emphasis in the Israeli-leaning environment. Symmetrically, a positive value (IS-skew) means the Israeli-leaning environment attributes the object comparatively more to Israeli actors. Further details are provided in \textit{Materials and Methods}. Robustness checks for different actor groupings, transcript segment lengths, and results for other actantial roles are included in the Appendix  (see Figures \ref{fig:polarization_robustness_actors}, \ref{fig:polarization_robustness_transcripts}, and \ref{fig:polarization_2}).

\section{Results}\label{sec:results}

\begin{figure*}
    \centering
    \includegraphics[width=1\linewidth]{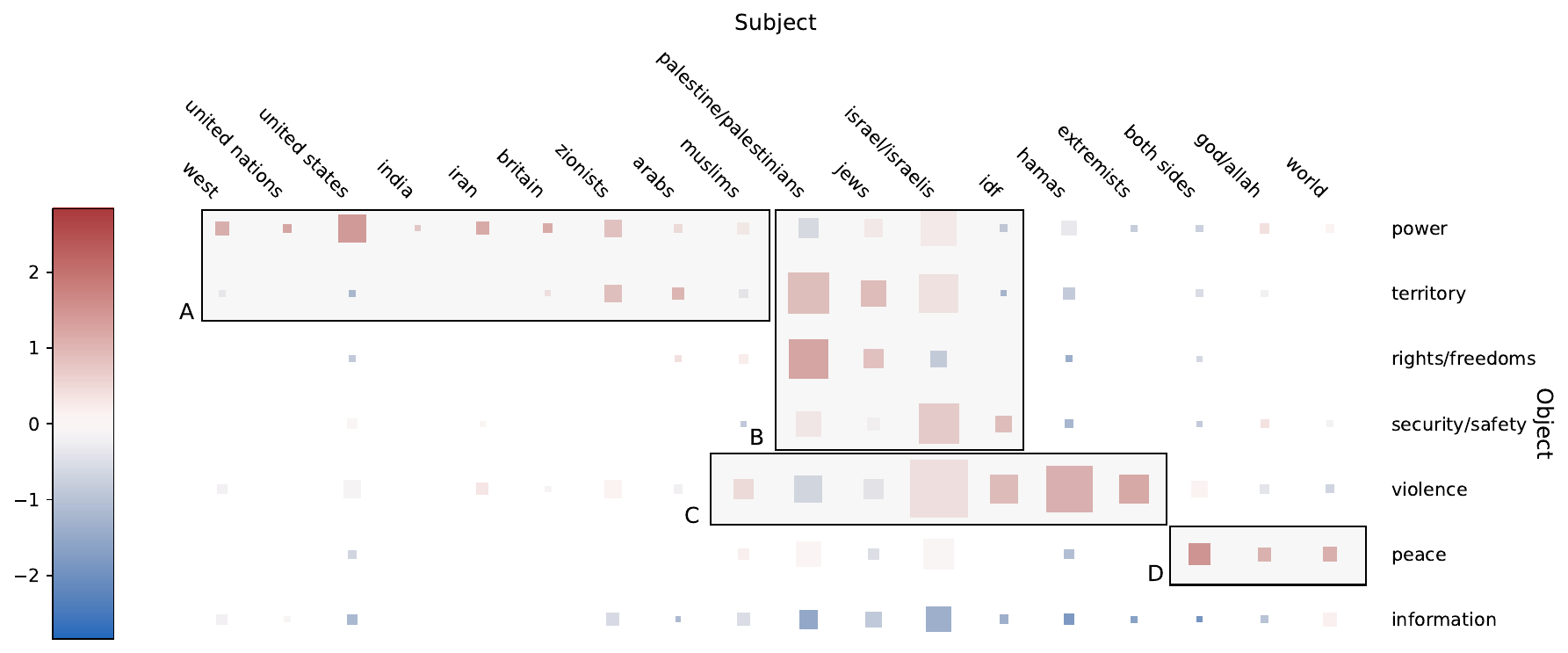}
    \caption{\textbf{Subject–Object co-occurrence patterns.} Each square represents the frequency of a given Subject (x-axis) and Object (y-axis) combination in the full dataset (n = 90,029 comments). Square size corresponds to the relative prevalence of each combination, with the most frequent being Israel/Israelis–violence (n = 6,078). Combinations with fewer than 60 comments are omitted for clarity. Comments referring to the meta-actors \textit{audience/commenter} or \textit{video creator} as Subject (28\%) are excluded. Color indicates the log ratio between observed and expected counts under the assumption of independence: positive values (red) denote combinations occurring more often than expected, and negative values (blue) denote combinations occurring less often.}
    \label{fig:heatmap}
\end{figure*}

The central axis of Greimas' model, comprising the Subject and the Object, constitutes what we refer to as a \textit{core narrative}. Unlike simple topic identification, this axis captures a directional relationship of desire, revealing whose interests center the discourse. Figure \ref{fig:heatmap} summarizes all Subject-Object combinations from our data across both query groups, thus providing a general overview of the discourse around the Israeli-Palestinian conflict on English-speaking YouTube. Region A includes external actors such as the United States or Iran, which are primarily focused on power interests. Most comments concentrate in regions B and C. Region B includes the actors Palestine/Palestinians, Jews, and Israel/Israelis, their territorial claims, and power interests. By contrast, the discourse around rights and freedoms mainly revolves around Palestinians, while security and safety are mostly discussed as Israeli interests. Region C identifies these same actors, as well as the IDF, Hamas, and extremists, as proponents of violence, with the main focus on Israel and Hamas. Region D covers appeals to peace by both sides, God/Allah, and the world. Comments focused on the meta-actors—audience/commenter and the video creator—were excluded from the visualization. These comments mainly focus on information as the Object and discuss the quality and accuracy of the video.

The general distribution shown in Figure \ref{fig:heatmap} is in line with common narratives around the conflict. In the following, we analyze how these patterns differ between the two information environments and between videos and comments. To this end, we calculate the shared probability mass, or overlap coefficient, between two distributions. This metric indicates how aligned two environments are in terms of their distribution of core narratives. A value of 1 indicates perfect alignment, while 0 indicates completely orthogonal discourse (see Equation \ref{eq:overlap}). Table \ref{tab:overlap_combined} reports the overlap coefficients between transcripts and comments within each environment, as well as between comments across environments and between transcripts across environments. Comments show moderate alignment with their respective videos, with overlap coefficients of 0.71 (Israeli-leaning) and 0.68 (Palestinian-leaning). Further, while transcripts across environments show lower alignment with 0.63, comments across environments show notably higher alignment with 0.8. This indicates that comments contain narratives that are less prominent in the videos, particularly narratives that are more evenly distributed across the two partisan information environments. 

\begin{table}
\caption{\textbf{Overlap coefficient.}
The overlap coefficient $O \in [0,1]$ is calculated on the probability distributions of Subject–Object combinations. Within-environment comparisons measure overlap between transcripts and comments within the same partisan environment ($O(T,C)$). Between-environments comparisons measure overlap between partisan environments for transcripts ($O(T,T)$) and comments ($O(C,C)$).}
\label{tab:overlap_combined}

\centering
\begin{tabular}{llc}
\toprule
Comparison type & Overlap coefficient \\
\midrule 
  Within Israeli-leaning environment & $O(T,C) = 0.71$ \\
  Within Palestinian-leaning environment & $O(T,C) = 0.68$ \\
\midrule
  Between Transcripts & $O(T,T) = 0.63$ \\
  Between Comments & $O(C,C) = 0.80$ \\
\bottomrule
\end{tabular}
\end{table}

Low overlap between the two environments, however, does not necessarily constitute polarization, that is, opposing representations of the conflict. For example, differences could be explained by a focus on British influence in the early 20th century in one environment, as opposed to US influence in the current conflict in the other. While this represents diverging discourse around the conflict, it merely constitutes a different issue focus rather than polarization.

\subsection{Subject divergence}\label{sec:results_narrative_polarization}
\begin{figure*}
    \centering
    \includegraphics[width=1\linewidth]{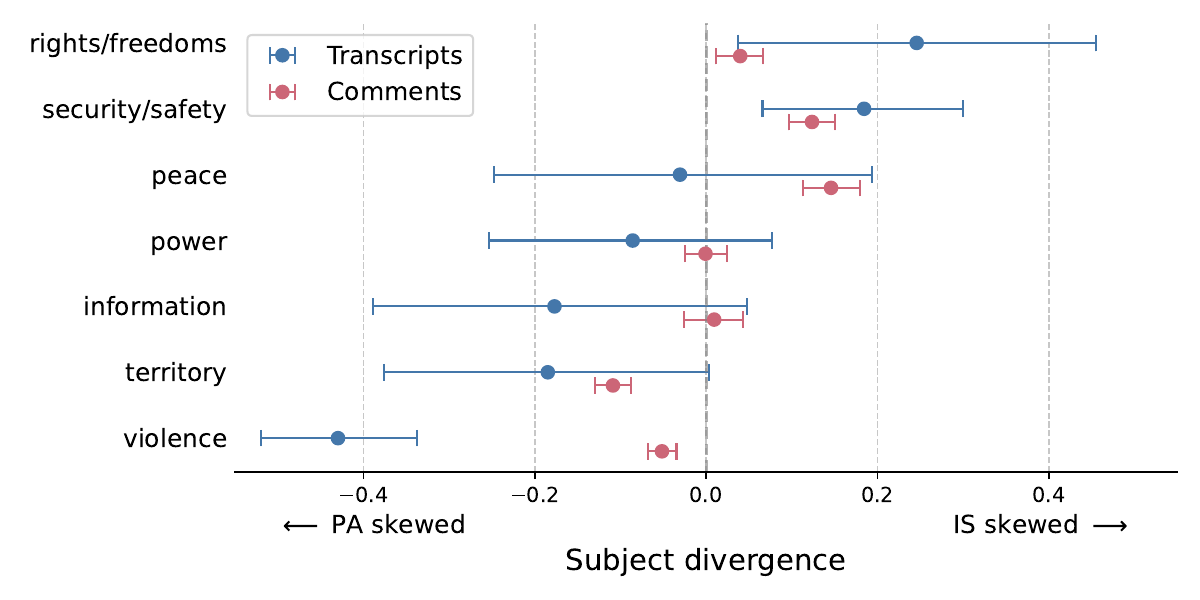}
    \caption{\textbf{Subject divergence in comments and transcripts.} Subject divergence refers to differences in how partisan information environments attribute the Subject role to different conflict actors. Shown are the attribution patterns for various Objects. Values near zero indicate low divergence, reflecting similar attribution patterns across environments. Negative values reflect greater attribution by the Israeli-leaning environment (relative to the Palestinian-leaning environment) to Palestinian actors, whereas positive values reflect greater attribution to Israeli actors. Error bars represent 95\% bootstrap confidence intervals (n=3000).}
    \label{fig:narrative_polarization}
\end{figure*}

To investigate whether core narrative overlap is linked to narrative polarization, we analyze the \textit{subject divergence} between the two partisan information environments.  To this end, we extract all transcripts and comments containing an Israeli or a Palestinian Subject, accounting for two-thirds of the data. Figure \ref{fig:narrative_polarization} reports subject divergence for all objects in both transcripts and comments. 

We find significant divergence across several of the seven objects for both transcripts and comments, confirmed using a permutation test (see Tables~\ref{tab:permutation_results_transcripts} and \ref{tab:permutation_results_comments} for details). The largest effect is observed in the transcripts for \textit{violence}, which is strongly PA-skewed. This means videos in the Israeli-leaning environment focus significantly more on Palestinians as the driving force of violence than videos in the Palestinian-leaning environment. Consequently, users in the two partisan environments are exposed to opposing representations of a central and contested aspect of the conflict. The pattern is reversed for \textit{security/safety} and \textit{rights/freedoms}, which are IS-skewed. The remaining objects are all PA-skewed, though these effects do not reach statistical significance.

The following examples illustrate how these actantial patterns manifest in the text. A video from the Israeli-leaning environment states: ``On October 7th, Hamas committed crimes against humanity. They raped, murdered, and violated Israeli women''~\cite{i24news_english_warning_2023}, positioning Hamas as the Subject, primarily associated with violence. By contrast, a video from the Palestinian-leaning environment assigns a similar role to Israel: ``Israel's war in Gaza has been one of the deadliest and most destructive in modern times''~\cite{channel_4_news_israeli_2024}. Additional examples from the comments are provided in Table \ref{tab:motif_examples}.

Comments generally follow the divergence direction of the transcripts. However, the effect size is significantly smaller. While the average absolute subject divergence in the transcripts amounts to $0.19$, it drops to $0.07$ in the comments. The difference is largest for \textit{violence}, where the PA-skew drops from $-0.43$ to $-0.05$. This confirms the assumption that higher overlap in core narratives of the comments, demonstrated earlier, stems from lower polarization rather than merely different issue focus. Consequently, key actors and their interests are represented more evenly in the comments. 

Notably, this lower level of subject divergence does not reflect reduced diversity of core narratives. We compare the probability distribution of all core narratives using normalized information entropy (see \textit{Materials and Methods}). We observe similar levels of entropy in transcripts ($0.84$, $95\%$ CI $[0.82, 0.85]$) and comments ($0.82$, $95\%$ CI $[0.82,0.83]$). Thus, comments are not more concentrated on a small set of dominant narratives, but exhibit lower divergence despite comparable narrative diversity. However, residual polarization remains: while the degree of divergence in the comments is lower, it remains significantly different from zero for several objects. 

\textit{Peace} constitutes a notable exception. While the transcripts show no divergence for this Object, the comments are IS-skewed. Hence, comments in the Israeli-leaning environment focus more on Israeli actors' desire for peace, while comments in the Palestinian-leaning environment focus more on Palestinian actors. This pattern may be explained by the prevalence of the narrative that attributes peace as an interest of \textit{both sides} (see Figure \ref{fig:heatmap}). Therefore, comments appealing to the peace interests of the respective out-group may be subsumed under general calls for peace, leaving the comments slightly skewed towards in-group appeals.

The relative trends between the two partisan environments emerge from different constellations of the absolute attribution patterns shown in Figure~\ref{fig:polarization_shares}. For most Objects, the comments of both partisan environments converge, though this does not necessarily entail an equal split of attribution between Israeli and Palestinian actors. For example, violence is attributed approximately equally to both actors in Israeli-leaning videos, but strongly to Israeli actors in Palestinian-leaning videos. The resulting PA-skew of the Israeli-leaning environment is lower in the comments as they attribute violence more to Israel, and Palestinian-leaning comments attribute it less to Israel. As a result, the average attribution shifts slightly toward Israel. In contrast, \textit{rights/freedoms} is consistently attributed more to Palestinians in both transcripts and comments and across both partisan environments. However, the extent of this attribution varies between environments. The reverse pattern appears for \textit{security/safety}, which is generally identified as an Israeli interest.

The results show that the comment sections exhibit similar distributions of narrative structures across partisan environments. Although videos diverge starkly on the perpetration of violence or the struggle for rights, the comment spaces of both partisan environments converge on general attribution patterns. However, these core narratives interact with the remaining roles of the Actantial Model that further shape the representations of the conflict actors.

\subsection{Narrative motifs}
Beyond the central axis of desire linking the Subject and the Object, the narrative framework includes additional functions, chiefly the Sender, Receiver, and Opponent. Figure~\ref{fig:polarization_2} shows divergence results for these additional actants. However, meaning emerges not only from individual actantial assignments but from their relational patterns. These include \textit{actant syncretisms}, when one actor occupies multiple roles, and actor opposition, when two actors are positioned on opposing sides of the actantial model's three axes: desire, communication, and power (Fig. \ref{fig:actantial_model}). These constellations, which we term \textit{narrative motifs}, carry interpretive weight beyond their constituent parts. Hence, the positioning of actors in different narrative motifs affects narrative polarization beyond subject divergence in the core narratives. 

\begin{figure*}
    \centering
    \includegraphics[width=1\linewidth]{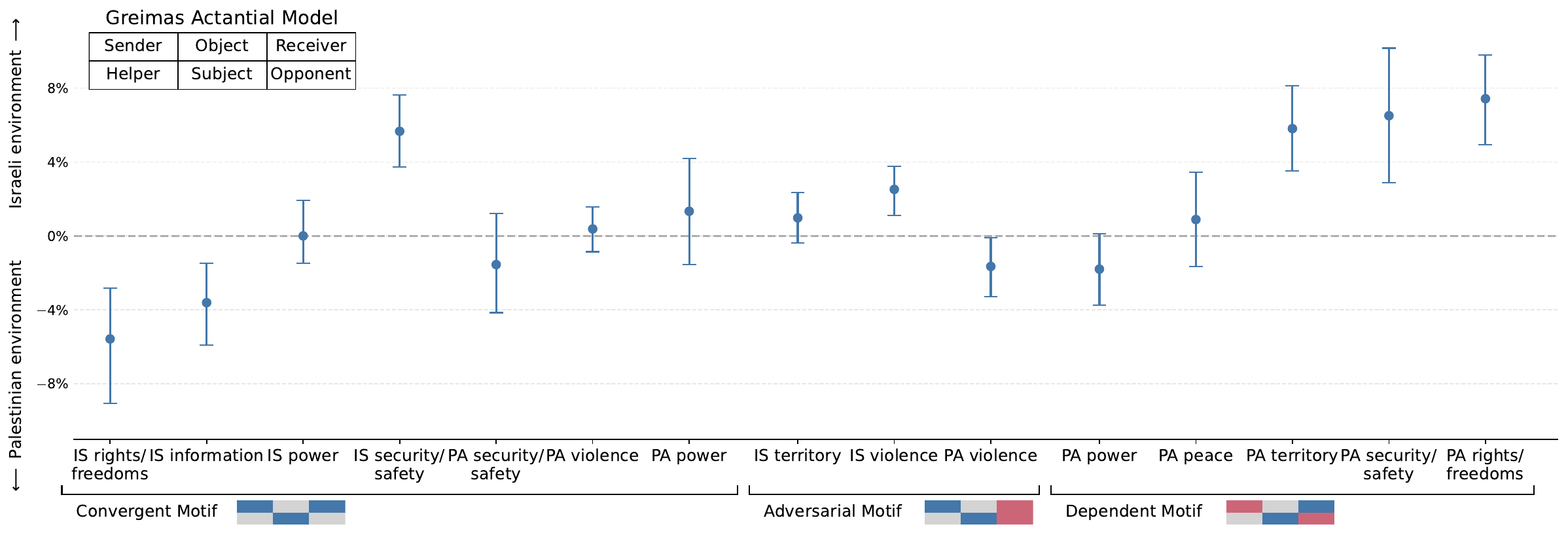}
    \caption{\textbf{Difference in narrative motif shares between partisan information environments.} Shown are three types of narrative motifs based on the Actantial Model (top left), with each representing a unique constellation of Israeli (IS) and Palestinian (PA) actors illustrated next to the motif titles. Markers indicate the relative motif difference between partisan environments for different core narratives. Motif actors are determined by the core narrative subjects (IS or PA) shown on the x-axis. Negative values indicate higher prevalence in the Palestinian-leaning environment, positive values indicate higher prevalence in the Israeli-leaning environment. Table \ref{tab:motif_examples} provides examples for each motif. Only narrative-motif combinations with an average prevalence above 5\% across partisan environments are shown. Error bars represent 95\% bootstrap confidence intervals (n=3000).}
    \label{fig:motif_attribution}
\end{figure*}

Three motifs prove particularly salient in our data: the convergent motif, the adversarial motif, and the dependent motif. The convergent motif positions the same actor as Subject, Sender, and Receiver. It focuses on the interests of one actor, its self-determination, but also internal conflict. The adversarial motif places actors in dual opposition. The Subject and Sender against Opponent and Receiver. The emphasis is on opposing interests, control through one actor, and the resistance of the other. The dependent motif inverts the Sender and Receiver. This creates a dynamic of dependency of the Subject and Receiver, and control through the Sender and Opponent. Figure \ref{fig:motif_attribution} showcases the difference between partisan information environments in the prevalence of these three motifs for various core narratives in our data. Table \ref{tab:motif_examples} provides examples for each of the motifs, and the absolute shares are shown in Figure~\ref{fig:narrative_motifs}.

\begin{table*}[!h]
\begin{adjustwidth}{-2.5cm}{-2.5cm}
\centering  
\caption{\textbf{Narrative motif examples.} Examples of comments illustrating the three salient narrative motifs shown in Figure \ref{fig:motif_attribution}: convergent, adversarial, and dependent. Each is defined by specific constellations of Israeli (IS) and Palestinian (PA) actors illustrated next to the motif titles. Each row shows a motif for a specific core narrative (Subject-Object) alongside a representative comment from the data. All examples are reproduced verbatim, preserving original spelling and grammar.}
\begin{tabularx}{\linewidth}{l X}
\toprule
\rowcolor{gray!20}
 \multicolumn{2}{l}{\textbf{Convergent Motif} \actantgrid{blue!60}{gray!60}{blue!60}{gray!60}{blue!60}{gray!60}}\\
\addlinespace[3pt]
    IS rights/freedoms & ``if a military is necessary.. those people serving, are still people, and then they should be taken care of.. not just be some pawns, for others to use and abuse.. [...]'' \\ 
    IS information & ``Zionism is a secte, a cult who brainwashed its members with a superiority complex, making them believe they are are special people with no accountability and the right to do all they want with non Jewish people. Israeli kids are growing in this cult completely brainwashed with atrocious lies about Palestinians!'' \\ 
    IS power & ``[...] Israeli ppl need to step up and save their country from Netanyahu before its too late'' \\ 
    IS security/safety & ``Complete and utter failure of Israeli intelligence and defence establishment. Military people in charge ,should be held responsible, and prosecuted to the full extend of the law.'' \\ 
    PA security/safety & ``They blame Israel… y dont they blame hamas for getting their homes destroyed. [...]'' \\
    PA violence & ``The people of Palestine donated and funded Hamas. [...] (Yes there are innocents there, I’m just saying that Palestine caused their own war.)'' \\ 
    PA power & ``Palestinians are the one who voted the Hamas into power. But look at what the Hamas did for Palestine ? [...]'' \\ 
    
\midrule
\rowcolor{gray!20}
\multicolumn{2}{l}{\textbf{Adversarial Motif} \actantgrid{blue!60}{gray!60}{red!60}{gray!60}{blue!60}{red!60}} \\
\addlinespace[3pt]
    IS territory & ``[...] It was never about the hostages nor about dismantling hamas  it's just meaningless revenge and a land grab'' \\ 
    IS violence & ``Israel drops bombs and fires missiles and launches artillery strikes against Palestinians killing thousands of civilians, but when those massacred people fight back and kill a few dozen of their oppressors suddenly it's ``terrorism" and ``war crimes" [...]'' \\ 
    PA violence & ``Hamas and their fighters are nothing but cowards. Women, children, elderly, and they’re proud of that. They killed the weak now it’s time for Israel. To settle this once and for all'' \\
    
\midrule
\rowcolor{gray!20}
\multicolumn{2}{l}{\textbf{Dependent Motif} \actantgrid{red!60}{gray!60}{blue!60}{gray!60}{blue!60}{red!60}} \\
\addlinespace[3pt]
     PA power & ``[...] the Palestinians is waging a war of independence from the cruel and destructive zionists , their influence over the Arabs has collapsed [...]" \\ 
     PA peace & ``There has been two peace agreements negotiated between Israel and Palestine. One with former Israeli Prime Minister Yitzhak Rabin, and then later on with Israeli Prime Minister Ehud Barack, but Netanyahu and Ariel Sharon decided not to implement them. [...]'' \\ 
     PA territory & ```The hostages are in gaza, and it’s wrong' says the beautiful occupier. You being there was wrong and still is wrong. Leave the land of the Palestinians.'' \\ 
     PA security/safety & ``[...] Gaza didn't receive a drop of aid of any kind for the first 46 days until thr first ceasefire. [...]'' \\
     PA rights/freedoms & ``[...] imagine being born there,living there, stuck in a place where the supply of food,water,electricity is cut off by an oppressing regime, where people have almost no freedom [...]'' \\ 
\bottomrule
\end{tabularx}
\label{tab:motif_examples}
\normalsize
\end{adjustwidth}
\end{table*}

The convergent motif appears in a range of core narratives, particularly when discussing Israeli desire for \textit{rights/freedoms} and \textit{security/safety} or Palestinian desire for \textit{power}. The Israeli narratives also show a significant difference between the environments, with security being more convergent in the Israeli-leaning discourse and rights/freedoms in the Palestinian-leaning one.  A close reading of a random sample of comments for each partisan environment and each narrative reveals the effects of this motif. The discourse around Israeli security interests focuses on how Israel provides security to its citizens. Notably, the Israeli-leaning environment is overwhelmingly critical, often discussing the failure of the Israeli government to provide adequate security measures to its citizens: ``Complete and utter failure of Israeli intelligence and defence establishment. Military people in charge ,should be held responsible, and prosecuted to the full extend of the law [sic].'' The security discourse outside of the convergent motif focuses less on this internal dimension and instead on common themes such as ``Israel's right to defend itself.'' Hence, the convergent motif reveals additional layers of actor complexity. Different parts of Israel are interacting with each other, in contrast to Israel being portrayed as a homogeneous entity in defense against an external threat. A similar trend is seen for rights and freedoms. The convergent motif is more prevalent in the Palestinian-leaning comments and surfaces tensions around Zionism and what it means for Jews around the world. 

The discourse around \textit{violence} from both Israeli and Palestinian actors is shaped by the adversarial motif. Here, comments emphasize the perpetration of violence from one side towards the other. The dual role of Receiver and Opponent often manifests itself in justifications of resistance or retaliation against the initially described violent act from the Subject and Sender. This pattern is notable given that Palestinian-leaning videos emphasize Israeli violence, yet some comments attempt to justify this violence by positioning it as a reaction to Palestinian violence under the adversarial motif: ``Hamas and their fighters are nothing but cowards. Women, children, elderly, and they’re proud of that. They killed the weak now it’s time for Israel. To settle this once and for all [sic].'' The inverse trend can be observed for Israeli-leaning videos.

The adversarial motif differs notably from the convergent motif, where violence is discussed as something that Palestinians had brought upon themselves: ``The people of Palestine donated and funded Hamas. [...] (Yes there are innocents there, I'm just saying that Palestine caused their own war.) [sic]''. In these examples, Israel is largely excluded from the discussion. Instead, Israeli violence is treated as a direct extension of Hamas' actions.

The dependent motif appears predominantly in narratives with Palestinian subjects. Following this motif, Palestinian \textit{rights/freedoms}, \textit{security/safety}, and \textit{territory} are both controlled by and denied through Israel. This emphasis is particularly strong under Israeli-leaning videos, where the motif is significantly more pronounced. While these points are also observed in the Palestinian-leaning discourse, the roles of Sender, Receiver, and Opponent are not always clearly specified together. Thus, for example, instead of a full dependent motif, comments only discuss one of these three relationships simultaneously. Consequently, the combined emphasis on control, entitlement, and suppression is relaxed in the Palestinian-leaning discourse.

These trends reveal how narrative polarization takes shape at different levels of complexity. Within core narratives, subject divergence affects general assumptions about the conflict, for example, defining the main perpetrator of violence. By incorporating additional actantial roles, narrative motifs reveal more nuanced actor qualities, including internal conflict, justification of violence, and dependency relationships. We have found that comments exhibit significantly lower levels of narrative polarization than the videos at the core narrative level. However, residual polarization persists at the motif level.

\section{Discussion}
We applied a novel framework for measuring polarization on a narrative dimension. We operationalized narrative through a structural framework of common character roles—actants—and introduced the notion of \textit{narrative polarization} to capture diverging representations of actors through their positioning in different narratives. We then investigated the narrative exposure of users in two partisan content environments. This revealed two important trends. First, videos from the two environments represented central actors in fundamentally different roles. Second, user comments exhibited significantly lower levels of narrative polarization, converging to a more even narrative distribution across partisan environments. This suggests that community discourse in online social media can expose users to less polarized narrative representations than provided in the videos. However, we found that rhetorical effects operate on various levels. Although users converged on core narratives involving only two actants, the discourse was conducive to a variety of more complex narrative motifs. These recurring constellations of multiple actantial roles revealed additional differences in the representation of central actors on a nuanced level. While the close reading examples suggested that some of these differences arise in reaction to the video content, we could not test this mechanism directly, leaving it as an open question for future work.

This work showcases the relevance of narrative in polarization research. While previous research has shown that users tend to cluster in homogeneous groups with little cross communication~\cite{bessi_users_2016, flamino_political_2023}, our work reveals that this limited mobility of users does not apply to narratives. Despite polarized narrative exposure through videos, users' comments converge on common narrative patterns, offsetting video bias. Hence, comment sections of polarized content exhibit lower narrative polarization than the videos and user interactions suggest. In contrast, previous research had found ideological alignment of comments with partisan news stories~\cite{han_news_2023}. This apparent tension suggests that narrative might capture a textual dimension that is partially orthogonal to user ideology. However, this hypothesis remains to be tested empirically.

At the same time, narrative and ideology interact through identity formation processes~\cite{somers_narrative_1994, polletta_deep_2017, bliuc_cooperation_2022}, meaning that content polarization and residual comment polarization may still shape users' long-term ideological positioning. Narrative motifs appear particularly conducive to such effects as they shape nuanced but important differences in the representation of central actors. For example, both content environments discuss Israeli security interests to similar extents, yet Israel is portrayed with greater internal complexity in Israeli-leaning discourse. Similarly, while violence is discussed at similar levels across environments, the narrative motifs surrounding justification diverge significantly. These patterns shape the perceived psychological depth of conflict actors. In-group actors are imbued with complex intentions, emotions, and motivations, while out-group actors receive more superficial, stereotypical representations.

Our narrative framework combines the natural language processing capabilities of large language models with rigorous narrative theory, yielding a framework that can infer complex actor relationships within a theoretically grounded and comprehensive structure. This advances both computational narrative analysis and narrative theory. We demonstrate the application of the Actantial Model at scale, extending its use beyond folktales and myths to large-scale computational analysis of social media discourse. The framework is readily applicable to textual data, requiring only partisan environment labels, a comprehensive actor label set, and human validation of the LLM performance. The consistent, actor-focused framework of the Actantial Model provides a narrative representation that can be applied to various topics and sources to investigate narrative polarization dynamics.

The broad applicability of this framework allows future research to move beyond some limitations of this study. Specifically, it is essential to assess whether the reduction of polarization in user comments holds across different issues and platforms. Further, we have analyzed the general discourse around an issue without accounting for popularity metrics, which have been found to amplify highly polarized content~\cite{robertson_users_2023, milli_engagement_2025}. Hence, depolarization in the aggregate distribution of comments may not reflect a user's actual experiences. Additionally, YouTube API search results have been shown to vary significantly over time~\cite{rieder_forgetful_2025}, suggesting value in replicating this analysis across different time points to assess both the stability of partisan information environments and the role of algorithmic curation in shaping polarization dynamics. Finally, analyzing narratives quantitatively inevitably involves trade-offs between interpretive depth and scale. While our framework demonstrates nuanced quantitative insights into discourse through narrative motifs, we compromised on actor label specificity to ensure consistency. This limits our ability to investigate finer-grained narrative differences, such as distinctions between different segments of Israeli and Palestinian societies.

Future work should address these limitations by examining narrative polarization across platforms and issues, investigating how narratives move across different media environments. This would not only test whether the depolarization observed in YouTube comments generalizes to other platforms, but also illuminate how narratives are collectively shaped across different societal spheres. Such research is crucial for identifying where narratives diverge and where online discourse fragments into incompatible representations of a shared reality.

\section{Materials and Methods}\label{sec:methodology}
\subsection{Query selection}
The labels pro-Palestinian and pro-Israeli are often used as shorthand but encompass a wide spectrum of views. Thus, to apply the concept of partisanship, we do not rely on a rigid definition of the two positions. Instead, we operationalize \textit{partisan search intent} via issues emphasized by the supporters of each side. In particular, we choose issues from pro-Palestinian and pro-Israeli protests in the United States. Previous research has shown that offline protests tend to coincide with heightened online attention~\cite{bastos_tents_2015}, motivating the use of offline protest issues as seeds to track online discussion. 

We rely on data from the Crowd Counting Consortium\footnote{\url{https://countingcrowds.org/2024/02/26/israel-palestine-protest-data-dashboards/}}. Their database of \textit{political crowds}, such as marches, protests, or strikes in the United States, includes more than 2700 pro-Israeli and 20900 pro-Palestinian events in the United States since October 7, 2023 until December 1, 2024. Each event is reported with a summary of the general claims of the respective crowd. These are a mix of annotator labels and verbatim quotes from banners and chants. We collect the most prominent issues from each group and identify related queries via Google Trends\footnote{\url{https://trends.google.com/trends/}}. We use variations of the popular query ``israel palestine explained'', to establish a baseline for search traffic and popular videos on the conflict. 

The final queries are listed in Table \ref{tab:queries}. Each set of queries represents a mix of issues related to the respective protest movement. Thus, partisan search intent does not necessarily capture a user's political leaning, but an awareness and interest in issues with high valence in the discourse of one of the two factions.

\begin{table}
\centering  
\caption{\textbf{Search queries.} Set of queries used to retrieve YouTube videos related to the Israeli–Palestinian conflict. Relative interest reflects the global number of searches for each query on YouTube between October 7, 2023 and October 1, 2024, expressed relative to the baseline query ``israel palestine explained''.}
\begin{tabular}{llr}
\toprule
Position & Query & Relative Interest\\
\midrule
\multirow{3}{*}{neutral} & ``israel palestine explained'' & $1.00$ \\
 & ``israel hamas explained'' & $0.20$ \\
 & ``israel gaza explained'' & $0.07$ \\
 \midrule
\multirow{6}{*}{pro-Israel} & ``attack on israel news'' & $1.22$ \\
 & ``israel october 7'' & $1.00$ \\
 & ``hamas terror'' & $0.07$ \\
 & ``hamas rape'' & $0.07$ \\
 & ``israel hostages'' & $0.67$ \\
 & ``israel antisemitism'' & $0.07$ \\
 \midrule
  \multirow{6}{*}{pro-Palestine} & ``attack on gaza news'' & $0.20$ \\
 & ``gaza rafah news'' & $0.33$ \\
 & ``gaza ceasefire'' & $0.93$ \\
 & ``gaza genocide'' & $0.67$ \\
 & ``israel war crimes'' & $0.20$ \\
 & ``israel zionism'' & $0.13$ \\
\bottomrule
\end{tabular}
\label{tab:queries}
\normalsize
\end{table}

\subsection{Data}\label{sec:data}
Data was collected using the YouTube Data API\footnote{\url{https://developers.google.com/youtube/v3/}}. The data was limited to videos of medium length, 4-20 minutes, English language, and published between October 7, 2023, and October 1, 2024. The lower time bound coincides with the Hamas terror attack on Israel, while the upper bound excludes the Israeli ground invasion into Lebanon starting on October 1, 2024. This time period was selected to create a more homogeneous topic space, focused on the conflict between Israel and Hamas in Gaza.

We collected the top 20 videos for each query in Table \ref{tab:queries}, ordered by relevance, yielding a total of 300 videos with 16\% duplicates. We dropped within-group duplicates, 39 videos with fewer than 20 comments, and one video duplicated across the two partisan content environments. Because the analysis focuses on exposure to narratives through partisan search terms, we also removed seven videos that appeared in both the neutral queries and at least one partisan query. The remaining set comprised 107 Israeli-leaning and 105 Palestinian-leaning videos. From each video, we collected all top-level comments, excluding replies. The final dataset was restricted to comments with a word count between 25 and 600, corresponding to the 75th and 99.9th percentiles of all comments. The strict lower bound reflects earlier experiments showing that, below this threshold, missing actants increased substantially, hindering reliable narrative identification.

Table \ref{tab:video_stats} presents the final numbers of videos, channels, and comments for each query group under these constraints. The comment count is strongly correlated with view count ($\rho = 0.84$). Thus, the more popular a video is, the more it is represented in the comment data. Further, the low overlap between the Israeli-leaning and the Palestinian-leaning environments (four out of 240 videos) confirms the assumption of partisanship of the search queries.

\begin{table*}[]
\centering
\small
\caption{\textbf{Video statistics.} Information environment level statistics, including the number of videos, transcript segments for a 150-word split, and total comments per environment. Entropy is the normalized Shannon entropy of the comment distribution over the different videos. Word count is the average word count of all comments for that environment.}
\begin{tabular}{lrrrrr}
\toprule
    Environment & Videos & Segments & Comments & Entropy & Word count \\
\midrule
    Neutral & $25$ & $291$ & $31832$ & $0.74$ & $67$ \\
    Israeli-leaning & $107$ & $912$ & $55894$ & $0.79$ & $63$ \\
    Palestinian-leaning & $105$ & $981$ & $34135$ & $0.86$ & $66$ \\
\bottomrule
\end{tabular}
\label{tab:video_stats}
\end{table*}

We transcribed the videos with OpenAI’s Whisper (\textit{whisper-large-v3})\footnote{\url{https://pypi.org/project/openai-whisper/}}. In contrast to YouTube's closed captions, this provides reliable punctuation, allowing transcripts to be split into coherent segments. Videos often contain multiple distinct narratives. Thus, we divide transcripts into segments, split at the first sentence boundary after 150 words. This corresponds to a long comment (95th percentile). We additionally provide robustness checks using alternative segment lengths (see Figure~\ref{fig:polarization_robustness_transcripts}).

\subsection{LLM annotations and validation}\label{sec:validation}
We annotate each comment and transcript segment with the six roles of the actantial model using an LLM (\textit{DeepSeek-R1-Distill-Qwen-32B}) at temperature 0, an approach developed in previous work~\cite{elfes_mapping_2025}. While nearly every text is annotated with a Subject and an Object, the remaining actants are not always clearly articulated. For example, a comment might discuss violence by one actor but not mention any opposition. Thus, each of the actants—Sender, Receiver, and Opponent—appears in approximately half of the comments, though not necessarily in the same comments. Table~\ref{tab:actant_stats} reports the exact shares of missing actants in both transcripts and comments.  The code used for annotations is available as a Python package\footnote{\url{https://github.com/jelfes/actantial}}. The annotator codebook and the test data are available in an OSF repository\footnote{\url{https://osf.io/qbaeg/}}.

The labeling task is highly complex due to the interdependencies of individual actants and the number of potential labels. We therefore compiled a comprehensive set of labels covering most of the discourse in the data while avoiding overlap. To this end, we annotated a test set of 11K comments with the LLM without predefined labels, then summarized these labels and complemented missing actors through several rounds of deliberation with human annotators. The final set comprises 21 actors and 7 objects (see Figure \ref{fig:heatmap}).

We selected a stratified sample of $330$ comments for hand annotations. All selections were made uniformly at random. We selected 10 channels from the unique set of channels for each of the three main query categories: neutral, Israeli-leaning, and Palestinian-leaning. Next, we selected one video for each channel and eleven comments from each video. Two expert annotators labeled the data following instructions from an extensive codebook. Due to the low quality of some comments, annotators had the opportunity to skip comments if they were unintelligible to them because of language, grammar, or spelling. We kept examples that were coded by at least one annotator. The final test set comprised $292$ comments of which $72\%$ were coded by both annotators. However, not every comment was annotated with all six actants, consistent with the pattern observed in the full dataset. Across all annotators, including the LLM, most comments were attributed a Subject and an Object, while Sender, Receiver, and Opponent were identified in approximately one-third of the comments. We dropped the Helper actant from the analysis, as it rarely occurs in the types of narratives found in the comments. 

Table \ref{tab:metrics_results} reports the inter-coder agreements between the three annotators—two humans and one LLM. The average agreement across the five actants is $0.73$, measured via a micro F1 score. This is calculated by assuming one coder as the ground truth and micro-averaging across all possible classes. Although, to the best of our knowledge, the coding of Greimas' Actantial Model is unprecedented at this scale, this measure facilitates the comparison with previous work on related concepts. Notably, Finlayson~\cite{finlayson_propplearner_2017} annotated folktales using Vladimir Propp's character functions—a predecessor to the Actantial Model—achieving a similar F1 score of $0.70$. Besides agreement scores between human annotators, our approach exceeds previous LLM approaches to Propp's narrative framework. Gervás and Méndez~\cite{gervas_tagging_2024} reported $0.46$ precision and $0.34$ recall using an LLM-coder on the same folktales as Finlayson, and Stammbach et al.~\cite{stammbach_heroes_2022} reported an average accuracy of 64\% for coding a subset of Propp's characters with an LLM in newspaper articles.

\begin{table}[]
\centering
\small
\caption{\textbf{Validation metrics.} Krippendorff's $\mathbf{\alpha}$ and F1-scores (macro, weighted, and micro) for 292 comments and three coders, including two expert human coders and one LLM.}
\begin{tabular}{lcccc}
\toprule
 & F1 (micro) / Acc. & F1 (weighted) & F1 (macro) & $\alpha$ \\
\midrule
Subject & 0.59 & 0.59 & 0.43 & 0.51  \\
Object & 0.74 & 0.74 & 0.71 & 0.63  \\
Sender & 0.84 & 0.84 & 0.54 & 0.64  \\
Receiver & 0.80 & 0.80 & 0.65 & 0.50  \\
Opponent & 0.68 & 0.68 & 0.47 & 0.64  \\
\midrule
Avg. & 0.73 & 0.73 & 0.56 & 0.59 \\
\bottomrule
\end{tabular}
\label{tab:metrics_results}
\end{table}

We further calculated Krippendorff's alpha, a standard measure for multi-annotator agreement. The average score for the five actants is $0.59$. This lies below the commonly assumed threshold of $0.67$, indicating worse performance than suggested by the micro F1 score. This discrepancy is explained by the weight attributed to rare classes. When using a macro F1 score instead, the value is close to Krippendorff's alpha. The macro average calculates the F1 score for each of the 21 labels and 7 objects separately, subsequently averaging across those individual scores. The micro score calculates one unified F1 score for all classes together. As a result, poor performance on low-prevalence classes has only a small impact on the micro score, while significantly reducing the macro score and Krippendorff's alpha. This is undesirable for our case, as the test sample is not representative of every individual label.

Besides the impact of low-prevalence classes, two main trends explain the level of residual disagreement. First, the main driver lowering overall accuracy is a high number of comments where annotators disagree on the annotation level—that is, either the Israeli-Palestinian conflict or the video itself. This is reflected by a high overlap between the categories \textit{commenter/audience} and one of the two main actors \textit{israel/israelis} and \textit{palestine/palestinians}. This overlap is not easily resolved, as most YouTube comments reflect an amalgamation of the commenter's own interests and their depiction of the conflict. The second trend is an overlap between \textit{israel/israelis} and \textit{palestine/palestinians} for the Subject. This is often due to two conflicting narrative perspectives. For example, A commits violent acts from which B needs protection, which can result in both narratives: A desires \textit{violence} or B desires \textit{security/safety}. 

Overall, our in-depth analysis of the labels reveals that our method performs on par or better than comparable previous approaches. Most disagreements can be traced back to differences in interpretation, rather than mistakes of the LLM. This is also supported by the fact that inter-coder agreement between the two human coders and the LLM is very similar. 

\subsection{Overlap Coefficient}\label{apx:overlap}
The overlap coefficient is defined as

\begin{equation}\label{eq:overlap}
    O(P, Q) = \sum_{x \in \mathcal X} \min\{p(x), q(x)\},
\end{equation}

with $\mathcal X$ the collection of all possible core narratives in the data and $p(x) \in [0, 1]$ the share of narrative $x$ in discourse $P$, and similarly for $q(x)$. As a result, $O(P, Q) \in [0, 1]$.

\subsection{Subject divergence}\label{apx:narrative_polarization}
Subject divergence is defined as the difference in within-environment attribution skews. Let $p(i, s)$ and $q(i, s)$ denote the proportion of attributions of object $i$ by the Israeli-leaning and Palestinian-leaning environments, respectively, to actor side $s\in\{\mathrm{IS},\mathrm{PA}, \mathrm{UNK}\}$. IS includes Israeli actors (Israel/Israelis, IDF, Zionists, Jews), PA are Palestinian actors (Palestine/Palestinians, Hamas, Arabs, Muslims). UNK represents both other actors (e.g., US, UN, Iran, ...) and unknown actors (i.e., actors that were not specified in the text). As the subjects fall under one of those three actor groups, we get 
\[
p(i, \mathrm{IS})+p(i, \mathrm{PA})+p(i, \mathrm{UNK})= 1
\]
and similarly for $q(i, s)$. We then define the within-environment attribution skews as
\[
A^{\mathrm{IL}}(i) = p(i, \mathrm{IS}) - p(i, \mathrm{PA}),\qquad
A^{\mathrm{PL}}(i) = q(i, \mathrm{IS}) - q(i, \mathrm{PA}).
\]

IL and PL denote the Israeli-leaning and Palestinian-leaning environments, respectively. The measure lies in $[-1, 1]$.

Subject divergence for object $i$ is the difference in skews between environments:

\begin{align*}
    D(i) &= \frac{A^{\mathrm{IL}}(i) - A^{\mathrm{PL}}(i)}{2} \\
    &= \frac{\big[p(i, \mathrm{IS})-p(i, \mathrm{PA})\big] - \big[q(i, \mathrm{IS})-q(i, \mathrm{PA})\big]}{2},
\end{align*}

which also takes values in $[-1,1]$. Negative values indicate a PA-skew of the IL environment as compared to the PL environment. That is the IL environment emphasizes Palestinian actors more than the PL environment. Positive values indicate an IS-skew of the IL environment. 

For our analysis, we filter all UNK actors. From $p(i, \mathrm{UNK}) = 0$ it follows that $p(i, \mathrm{PA}) = 1 - p(i, \mathrm{IS})$ and the formula for subject divergence reduces to equation \ref{eq:sub_div_red} as shown in the results section:

\[
    D(i) = p(i, \mathrm{IS}) - q(i, \mathrm{IS}).
\]

\subsection{Narrative Entropy}

To assess the diversity of core narratives, we compute the normalized Shannon entropy of the narrative distribution. Let $P=\{p(x)\}_{x\in\mathcal{X}}$ denote the distribution of narratives in a given discourse, where $p(x)$ is the share of narrative $x\in\mathcal{X}$ and $\sum_{x\in\mathcal{X}} p(x) = 1$. Let $N=|\mathcal{X}|$ denote the number of unique core narratives. Normalized entropy is defined as

\[
    H(P) = -\frac{1}{\ln N}\sum_{x\in\mathcal{X}} p(x) \ln p(x),
\]

where $\ln$ denotes the natural logarithm, and division by $\ln N$ scales the measure to $[0, 1]$. $0$ indicates a fully concentrated distribution and $1$ a uniform distribution over all narratives. Confidence intervals are computed via 3,000 bootstrap resamples.

\bibliography{references}

\section{Acknowledgments}
We thank Aisha Nazim and our annotators for their contributions to developing the annotation guidelines and their careful annotation work.

\section{Competing interest}
The authors declare no competing interests.

\section{Funding}
This research has been jointly funded by Taighde Éireann – Research Ireland under grant number GOIPG/2023/4198, by the Danish Data Science Academy under grant number 2024-2725, by the National Council for Scientific and Technological Development under grant number 406504/2022-9, and by the Carlsberg Foundation through the COCOONS project under grant number CF21-0432.

\section{Author contributions statement}
J.E., M.B., and L.M.A. conceptualized the research study. J.E. collected and analyzed the data, created visualizations, and wrote the first draft of the manuscript. J.E., M.B., and L.M.A. revised the manuscript.

\clearpage
\begin{appendices}
\section{}

\renewcommand{\thefigure}{A\arabic{figure}}
\setcounter{figure}{0}

\renewcommand{\thetable}{A\arabic{table}}
\setcounter{table}{0}

\begin{figure}[!h]
\centering

\includegraphics[width=1\textwidth]{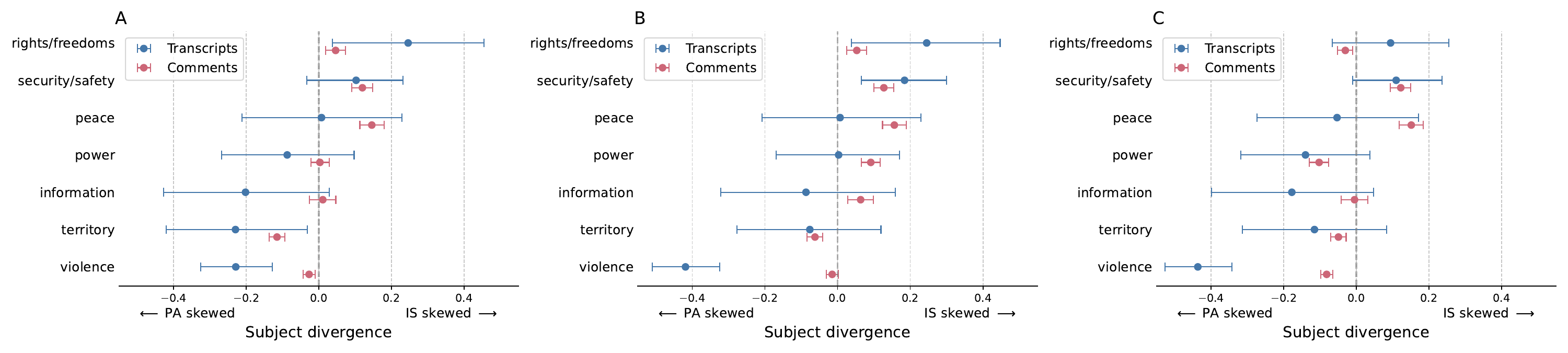}

\caption{\textbf{Subject divergence robustness check: actor groupings.} Shown are the attribution patterns of the Subject role for different actors, varying the actor groupings used in the main text. The main-text grouping (Fig.~3) categorizes Israeli actors as IDF, Israel/Israelis, Jews and Zionists, and Palestinian actors as Arabs, Hamas, Muslims and Palestine/Palestinians. Panels (A)–(C) vary these groupings by dropping actor pairs:
(A) IDF and Hamas dropped.
(B) Zionists and Arabs dropped.
(C) Jews and Muslims dropped.
Error bars represent 95\% bootstrap confidence intervals (n=3000).}
\label{fig:polarization_robustness_actors}
\end{figure}

\begin{figure}
    \centering
    \includegraphics[width=1\linewidth]{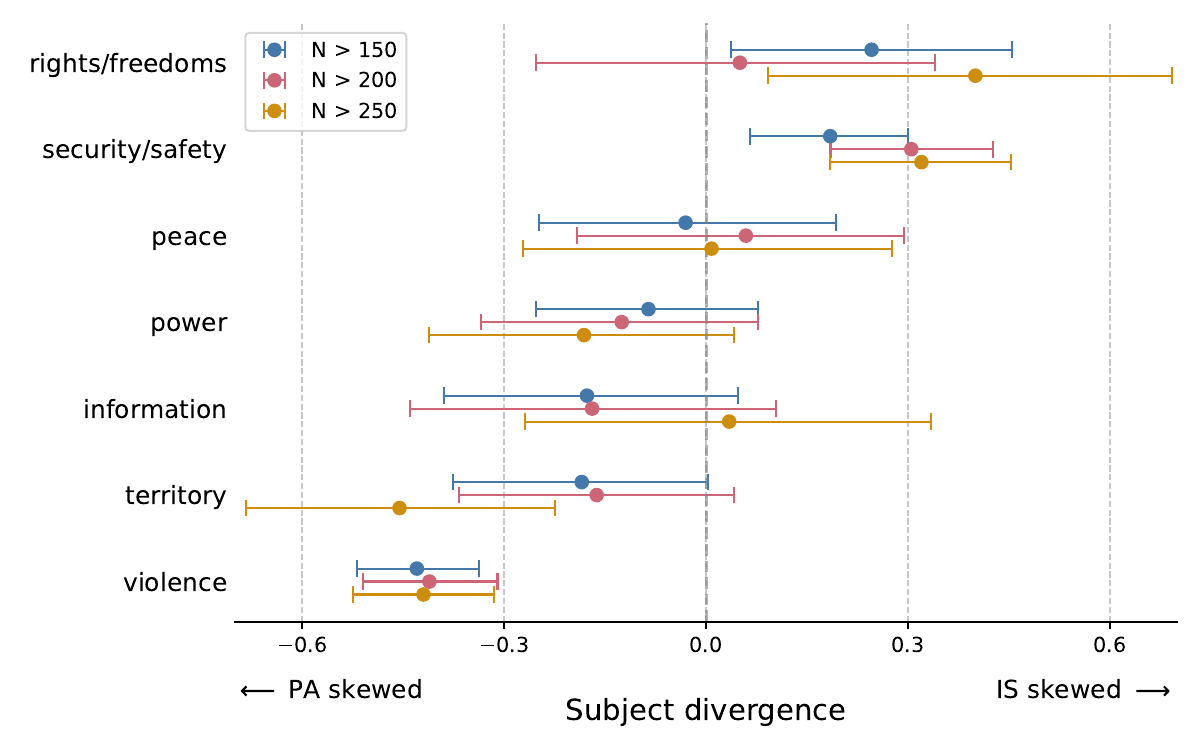}
    
    \caption{\textbf{Subject divergence robustness test: transcript length.} Shown are the attribution patterns of the Subject role for different actors. The figure mirrors the main-text analysis (Fig.~3), but varies the minimum length of the transcript segments from 150 in the main to 250.
    Error bars represent 95\% bootstrap confidence intervals (n=3000).}
    \label{fig:polarization_robustness_transcripts}
\end{figure}

\begin{figure*}
    \centering
        \centering        
    \includegraphics[width=1\linewidth]{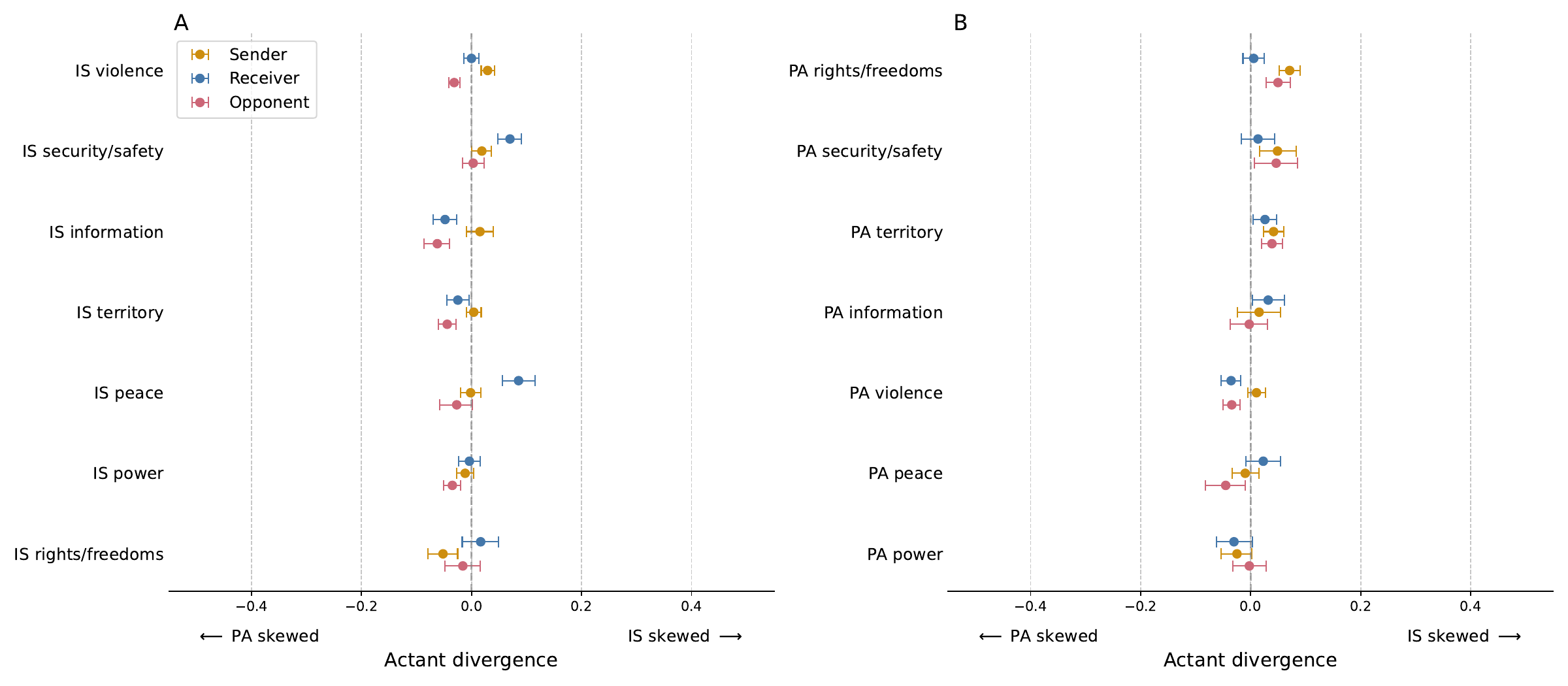}
    \caption{\textbf{Actant divergence in the comments.} Shown are the attribution patterns of the Sender, Receiver, and Opponent roles across Subject–Object combinations with an Israeli Subject (A) or a Palestinian Subject (B). Negative values reflect greater attribution of a role by the Israeli-leaning environment (relative to the Palestinian-leaning environment) to Palestinian actors, whereas positive values reflect greater attribution to Israeli actors. Error bars represent 95\% bootstrap confidence intervals (n=3000).}
    \label{fig:polarization_2}
\end{figure*}

\begin{figure*}
\centering

\includegraphics[width=1\textwidth]{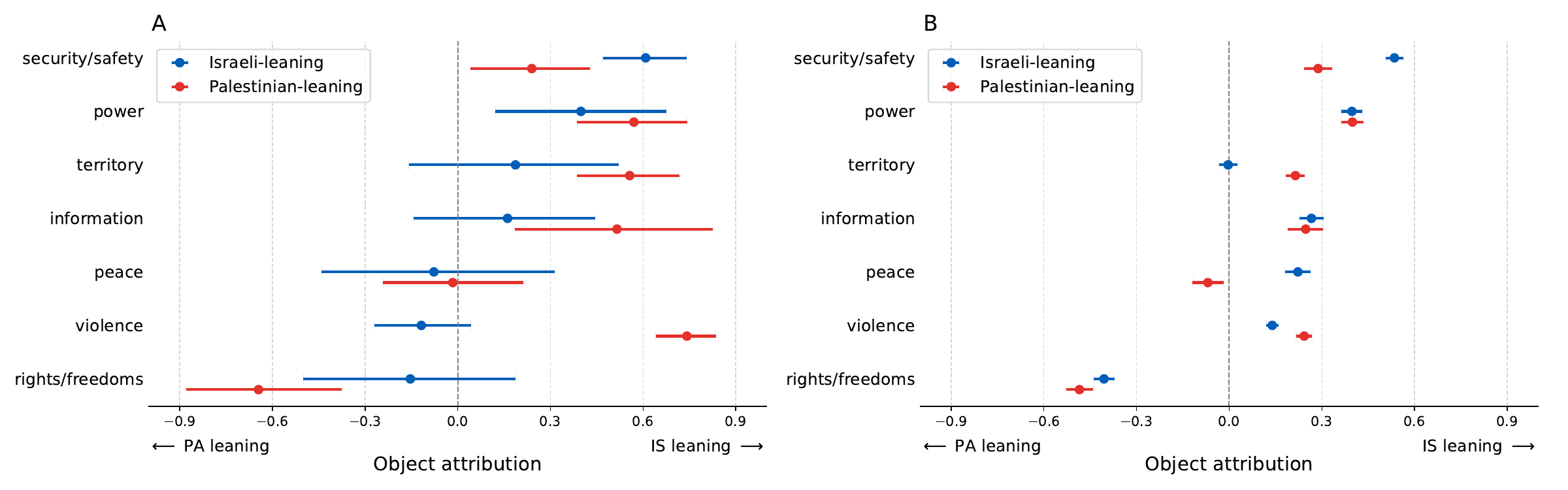}

\caption{\textbf{Object attribution in transcripts (A) and comments (B).}
Object attribution captures the degree to which Palestinian-leaning (red) and Israeli-leaning (blue) environments attribute objects to conflict actors. Values near zero indicate equal attribution to both Israeli and Palestinian actors. Values close to $-1$ indicate strong attribution to Palestinian actors, while values close to $1$ indicate strong attribution to Israeli actors. Error bars represent 95\% bootstrap confidence intervals ($n=3000$).}
\label{fig:polarization_shares}
\end{figure*}

\begin{figure}[h!]
    \centering
    \includegraphics[width=1\linewidth]{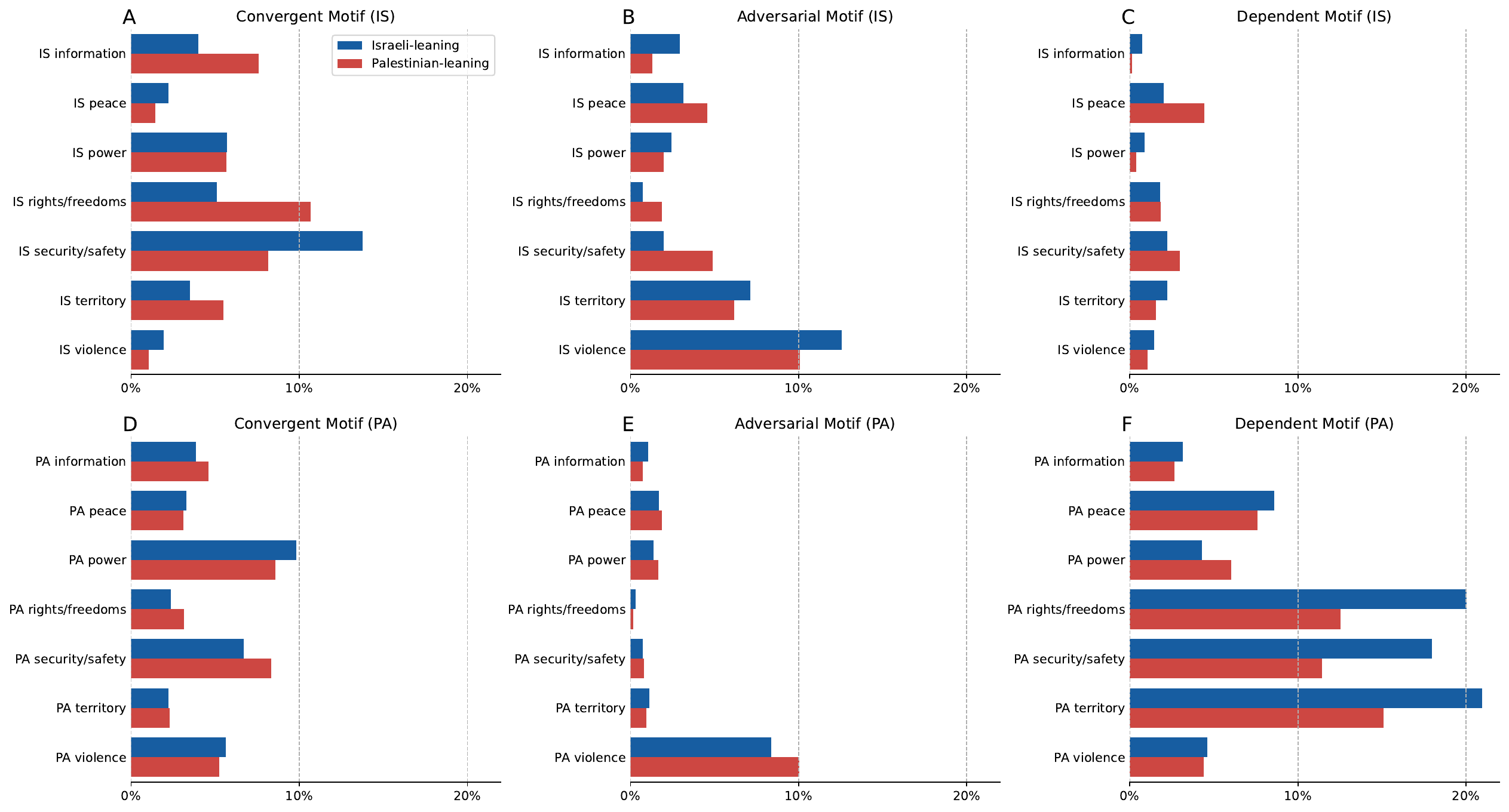}
    \caption{\textbf{Narrative motifs for each core narrative in the comments.} Values indicate the share of a specific motif for each core narrative within a partisan environment. The top row (A–C) focuses on Israeli actors as the Subject, the bottom row (D–F) on Palestinian actors. The convergent motif positions one side simultaneously as Subject, Sender, and Receiver. The adversarial motif positions one side as Subject and Sender, the other as Receiver and Opponent. The dependent motif positions one side as Subject and Receiver, the other as Sender and Opponent.}
    \label{fig:narrative_motifs}
\end{figure}

\clearpage

\begin{table}
\centering
\caption{\textbf{Permutation test transcripts.} Observed divergence shows the subject divergence between the partisan environments reported in Figure~3. \textit{p}-values are computed using a permutation test (n=3000). Bonferroni column reports adjusted \textit{p}-values for multiple comparisons.}
\label{tab:permutation_results_transcripts}
\begin{tabular}{lccc}
\toprule
Object & Observed divergence & \textit{p}-value & Bonferroni \\
\midrule
rights/freedoms & $0.248$* & 0.030 & 0.210 \\
violence & $-0.430$*** & $<0.001$ & $<0.001$ \\
peace & $-0.029$ & 0.766 & $1.000$ \\
information & $-0.177$ & 0.120 & 0.842 \\
territory & $-0.184$* & 0.020 & 0.142 \\
power & $-0.087$ & 0.290 & $1.000$ \\
security/safety & $0.183$** & 0.002 & 0.016 \\

\bottomrule
\end{tabular}
\footnotesize
*** \textit{p}$<0.001$, ** \textit{p}$<0.01$, * \textit{p}$<0.05$
\end{table}

\begin{table}[h!]
\centering
\caption{\textbf{Permutation test comments.} Observed divergence shows the subject divergence between the partisan environments reported in Figure~3. \textit{p}-values are computed using a permutation test (n=3000). Bonferroni column reports adjusted \textit{p}-values for multiple comparisons.}
\label{tab:permutation_results_comments}
\begin{tabular}{lccc}
\toprule
Object & Observed divergence & \textit{p}-value & Bonferroni \\
\midrule
rights/freedoms & $0.040$** & 0.004 & 0.028 \\
territory & $-0.109$*** & $<0.001$ & $<0.001$ \\
violence & $-0.051$*** & $<0.001$ & $<0.001$ \\
peace & $0.146$*** & $<0.001$ & $<0.001$ \\
information & $0.010$ & 0.561 & $1.000$ \\
power & $-0.000$ & 0.973 & $1.000$ \\
security/safety & $0.123$*** & $<0.001$ & $<0.001$ \\
\bottomrule
\end{tabular}
\footnotesize

*** \textit{p}$<0.001$, ** \textit{p}$<0.01$, * \textit{p}$<0.05$
\end{table}

\begin{table}[h!]
\centering
\footnotesize
\caption{\textbf{Share of missing actants in the transcripts and the comments.}}
\begin{tabular}{lrr}
\toprule
 Actant & Transcripts & Comments \\
\midrule

Subject & 9\% & 8\%\\
Object & 1\% & 0\%\\
Sender & 43\% & 44\%\\
Receiver & 36\% & 38\%\\
Helper & 74\% & 83\%\\
Opponent & 45\% & 53\%\\
\bottomrule
\end{tabular}
\label{tab:actant_stats}
\end{table}

\end{appendices}

\end{document}